\documentclass[prb, aps, twocolumn, superscriptaddress]{revtex4}
\usepackage{graphicx}
\usepackage{bm}
\usepackage{amssymb}
\usepackage{amsmath}
\usepackage{amsfonts}

\newcommand* {\vek}[1]{\bm{#1}}
\newcommand* {\kk}{\bm{k}}
\newcommand* {\rr}{\bm{r}}
\newcommand* {\expect}[1]{\langle #1\rangle}
\newcommand* {\frack}[2]{{\textstyle\frac{#1}{#2}}}
\newcommand{\marr}[2][c]{\left({\renewcommand{\arraystretch}{0.8}%
  \tabcolsep 0pt\begin{array}{@{}#1@{}} #2 \end{array}}\right)}

\begin{document}
\title{Spin angular impulse due to spin-dependent reflection off a
barrier}

\author{V. Teodorescu}
\affiliation{Department of Physics, Northern Illinois University,
DeKalb, IL 60115}
\author{R. Winkler}
\affiliation{Department of Physics, Northern Illinois University,
DeKalb, IL 60115}
\affiliation{Materials Science Division, Argonne National
Laboratory, Argonne, IL 60439}

\begin{abstract}
  The spin-dependent elastic reflection of quasi two-dimensional
  electrons from a lateral impenetrable barrier in the presence of
  band-structure spin-orbit coupling results in a spin angular
  impulse exerted on the electrons which is proportional to the
  nontrivial difference between the electrons' momentum and
  velocity. Even for an unpolarized incoming beam we find that the
  spin angular impulse is nonzero when averaged over all components
  of the reflected beam. We present a detailed analysis of the
  kinematics of this process.
\end{abstract}
\date{July 23, 2009}
\maketitle

Spin-dependent scattering in confined systems with spin-orbit
coupling (SOC) offers fascinating possibilities to manipulate the
electrons' spin degree of freedom if the electrons move through an
appropriate orbital environment. In the presence of band-structure
SOC the elastic reflection of quasi-two-dimensional (2D) electrons
from a lateral impenetrable barrier depends on their spin
orientation so that such a setup can act like a spin filter.
\cite{che05} Also, one can obtain spin accumulation near the
barrier. \cite{usa05} Scattering off circular barriers was
investigated in Refs.~\onlinecite{yeh06, wal06}. Several groups
studied the propagation of electrons in systems where the magnitude
of SOC is modulated in space. \cite{kho04, ram04, pal06, che07,
 sanc06} A related configuration uses a magnetic field perpendicular
to the plane of a quasi-2D system which results in spin-dependent
magnetic focusing. \cite{usa04, rok04} The studies of these systems
focused on the spatial separation of the trajectories of electrons
with different spin orientations, on the generation of a spin
polarization, and on interference effects related to the different
paths that scattered electrons can take. These phenomena complement
the spin precession that characterizes the propagation of electrons
moving freely in the effective magnetic field characterizing SOC.
\cite{dat90, win03}

Here we show that scattering off barriers also allows one to
manipulate the spin degree of freedom in a conceptually different
way as SOC results in an effective spin torque that can change the
orientation of the spin vector nonadiabatically during the
scattering process. Recently, spin torques have been a subject of
significant interest as a tool for reorienting the magnetization
direction of magnetic layers. \cite{ral08z} Spin-dependent
scattering off barriers give rise to spin torques in a well-defined
setting. The effect is proportional to the nontrivial difference
between the electrons' momentum and velocity. When integrated over
the duration of the scattering process, the spin torque corresponds
to a spin angular impulse. Like the mechanical torque discussed by
Mal'shukov et al., \cite{mal05a} the spin angular impulse is a
manifestation of the fundamental conservation laws characterizing
the electron dynamics in the presence of SOC. We show that even for
an unpolarized incoming beam the spin angular impulse is nonzero
when averaged over all components of the reflected beam. The effect
is the largest in magnitude if the angle of the incoming beam
relative to the reflecting barrier approaches a critical value. Our
findings are relevant for a large variety of transport experiments
in confined geometries. \cite{zut04}

For this study, we consider the Hamiltonian
\begin{equation}
  \label{eq:ham}
  H = \frack{1}{2}\mu k^2 + \alpha (k_y \sigma_x  - k_x \sigma_y) + V(x),
\end{equation}
where $\kk = (k_x, k_y, 0)$ is the 2D in-plane wave vector and $\mu
\equiv \hbar^2 / m^\ast$ with the effective mass $m^\ast$. The
second term in Eq.\ (\ref{eq:ham}) is the Rashba SOC \cite{byc84}
with Rashba coefficient $\alpha > 0$, and $\sigma_i$ are the Pauli
spin matrices. Finally, $V(x)$ is the potential due to the
impenetrable barrier. We assume $V(x) = 0$ for $x<0$ and $V(x) =
\infty$ for $x>0$. Previous studies showed that smoother gradients
preserve the important physics. \cite{kho04, sil06} While we
restrict ourselves for conceptual clarity to Rashba SOC, it is
straightforward to include other contributions to SOC such as
Dresselhaus SOC. \cite{dre55a} The spin-split dispersion is
\begin{equation}
  \label{eq:disp}
  E_\pm (\kk) = \frack{1}{2}\mu k^2 \pm \alpha k.
\end{equation}
For a given density $N$ the dispersion (\ref{eq:disp}) results in
Fermi wave vectors \cite{win03}
\begin{equation}
  \label{eq:fermiwv}
  k_\pm = \sqrt{2\pi \Big(N \mp \frac{\alpha}{\pi \mu^2}
  \sqrt{2\pi \mu^2 N - \alpha^2} \Big)} 
  = \frac{1}{\mu} \big( \sqrt{2\mu E_F + \alpha^2} \mp \alpha\big),
\end{equation}
where $E_F = E_+ (\kk_+) = E_- (\kk_-)$ is the Fermi energy. We note
that $k_- - k_+ = 2\alpha / \mu > 0$ independent of $N$
(Ref.~\onlinecite{dat90}) provided that $k_- > 2\alpha / \mu$, i.e.,
$N > N_q \equiv \alpha^2 / (\pi\mu^2)$). As $N$ is typically much
larger than the ``quantum density'' $N_q$, the case $N < N_q$ is
ignored in the following. \cite{quantdens} Yet we note that all
formulas developed below give the largest observable effects for
small densities, consistent with the fact that the SOC term in Eq.\
(\ref{eq:ham}) is most important for small densities.

\begin{figure}
\begin{center}
\includegraphics[width=0.98\columnwidth]{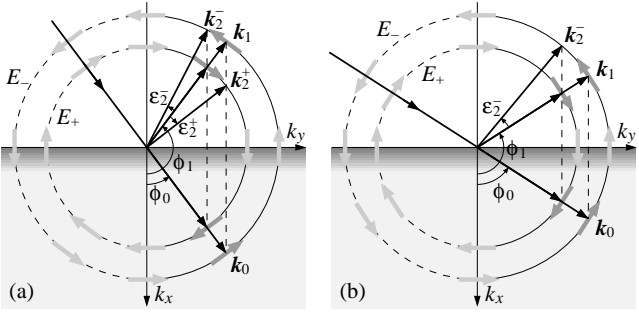}
\caption{\label{fig:sketch} Sketch illustrating the spin-dependent
reflection from a barrier at $x=0$ in the (a) undercritical and (b)
overcritical regime. The circles show the Fermi contours $E_\pm$ in
the $(k_x, k_y)$ plane. The gray arrows indicate the spin
orientation of the eigenstates along the Fermi contours. $\kk_0$,
$\kk_1$, and $\kk_2^\mp$ are the wave vectors of the incoming, the
ordinarily reflected, and the extraordinarily reflected beams. (We
omit the subscripts $\pm$ of $\kk_0$ and $\kk_1$.)}
\end{center}
\end{figure}

\begin{figure}[t]
\begin{center}
\includegraphics[width=0.70\columnwidth]{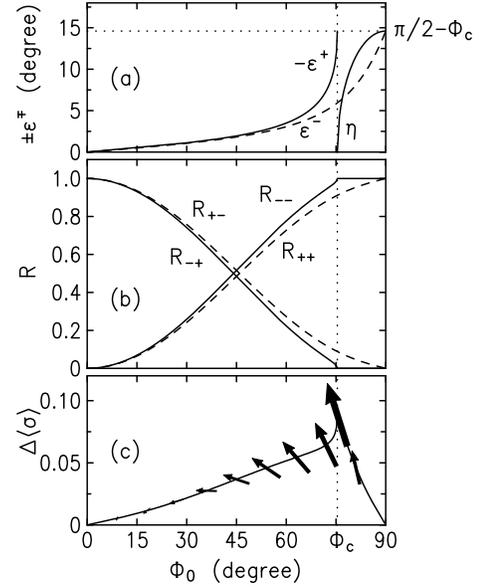}
\caption{\label{fig:results} (a) Splitting angles $\epsilon^\mp$ and
 out-of-plane spin orientation $\eta$, (b) reflection coefficients
 $R_{\pm\pm}$ and (c) average spin angular impulse $\Delta
 \expect{\sigma}$ (assuming an unpolarized incoming beam) as a
 function of the angle $\phi_0$ of the incoming beam for an InSb 2D
 electron system with density $N=2\times 10^{11}$~cm$^{-2}$,
 effective mass $m^\ast = 0.014~m_0$, and Rashba coefficient $\alpha
 = 0.1$~eV$\:$\AA. The critical angle $\phi_c = 75.4^\circ$ is
 marked by a dotted vertical line. The arrows in (c) indicate the
 orientation of $\Delta \expect{\vek{\sigma}}$ for the coordinate
 system in Fig.~\ref{fig:sketch}.}
\end{center}
\end{figure}

We consider a ballistic electron beam with wave vector $\kk_0^\pm$
that is reflected elastically from the barrier at $x=0$, see
Fig.~\ref{fig:sketch}. The wave functions are \cite{win03}
\begin{equation}
  \label{eq:wave}
  \begin{array}[b]{rl}
  \psi_{\kk_\pm} (\rr) = & \displaystyle
  \frac{A_0 \, e^{i \kk_0^\pm \cdot \rr}}{\sqrt{2}}
  \marr{1 \\ \mp ie^{i\phi_0}}
  + \frac{A_1^\pm \, e^{i \kk_1^\pm \cdot \rr}}{\sqrt{2}}
  \marr{1 \\ \mp ie^{i\phi_1}}
  \hspace{-2em} \\[2.5ex] & \displaystyle
  + \frac{A_2^\mp \, e^{i \kk_2^\mp \cdot \rr}}{\sqrt{2}}
  \marr{1 \\ \pm ie^{i\phi_2^\mp}},
\end{array}
\end{equation}
where $\phi_i$ is the polar angle of $\kk_i^\pm$. Throughout, the
index $0$ refers to the incoming beams, $1$ ($2$) denotes the
ordinarily (extraordinarily) reflected beam preserving (not
preserving) the magnitude of $k$, and the indices $\pm$ are defined
via Eq.\ (\ref{eq:fermiwv}). Translational invariance parallel to
the barrier $x=0$ implies the conservation of the $y$ component of
crystal momentum, i.e.,
\begin{equation}
  \label{eq:ky}
  k_y^\pm = k_\pm \sin\phi_0 = k_\pm \sin\phi_1 = k_\mp \sin\phi_2^\mp,
\end{equation}
which yields the ordinary reflection law
\begin{equation}
  \label{eq:ord}
  \phi_1 = \pi - \phi_0,
\end{equation}
and the extraordinary reflection law
\begin{equation}
  \label{eq:extraord}
  \phi_2^\mp = \pi - \arcsin \left(\frac{k_\pm}{k_\mp}\: \sin{\phi_0} \right).
\end{equation}
As $k_- > k_+$, the equation for $\phi_2^-$ has a real solution for
any $0 \le \phi_0 \le \pi/2$. However, a real solution for $\phi_2^+$
exists only for $0 \le \phi_0 \le \phi_c$, where
\begin{equation}
  \label{eq:phic}
  \phi_c \equiv \arcsin (k_+/k_-).
\end{equation}
For $\phi_0 > \phi_c$, Eq.\ (\ref{eq:extraord}) becomes equivalent
to
\begin{equation}
  \label{eq:phicomp}
  \phi_2^+ = \frac{\pi}{2} + i \ln \bigg[ \frac{\sin \phi_0}{\sin \phi_c}
    + \sqrt{\bigg(\frac{\sin \phi_0}{\sin \phi_c}\bigg)^2 - 1} \bigg].
\end{equation}
It will become evident below that the critical angle $\phi_c$ plays
an important role for many geometrical aspects of this problem.
Note that $\phi_c \rightarrow 0$ for $N \rightarrow N_q$. The
dif\-ference between the angles of the two reflected beams is
\begin{equation}
  \label{eps}
  \epsilon_\mp \equiv \phi_2^\mp - \phi_1
  = \phi_0 - \arcsin \left(\frac{k_\pm}{k_\mp}\: \sin{\phi_0} \right).
\end{equation}
The splitting angle $\epsilon_-$ is positive and its largest value
is obtained for $\phi_0 \rightarrow \pi/2$ (grazing incidence)
giving
\begin{equation}
  \label{eq:epsmax}
  |\epsilon_\mathrm{max}|
  = \pi/2 - \phi_c.
\end{equation}
The angle $\epsilon_+$ is negative and its largest value in
magnitude is obtained for $\phi_0 = \phi_c$. Yet the corresponding
value $|\epsilon_\mathrm{max}|$ is again given by Eq.\
(\ref{eq:epsmax}), see Fig.\ \ref{fig:results}(a).

Conservation of the wave vector component $k_y^\pm$ implies that
$k_{x1}^\pm$ and $k_{x2}^\mp$ become functions of $\phi_1$ and
$\phi_2^\mp$
\begin{equation}
  \label{eq:kx}
  k_{x1}^\pm = k_\pm \cos\phi_1, \qquad
  k_{x2}^\mp = k_\mp \cos\phi_2^\mp.
\end{equation}
For a complex angle $\phi_2^+$ the wave vector $k_{x2}^+$ becomes
imaginary, \cite{usa05} i.e., $\kappa_2^+ \!\! \equiv ik_{x2}^+ > 0$
describes an exponentially decaying solution (for $x \le 0$).

Continuity of the wave function $\psi_{\kk_\pm} (\rr)$ at the
interface $x=0$ yields the conditions
\begin{equation}
  \label{eq:coeff}
  \frac{A_1^\pm}{A_0}
  = \frac{e^{2i\phi_0}-e^{i\epsilon_\mp}}{1+e^{i\epsilon_\mp}},
  \qquad
  \frac{A_2^\mp}{A_0}
  = -\frac{1+e^{2i\phi_0}}{1+e^{i\epsilon_\mp}}.
\end{equation}
Here the expression for $A_2^+$ refers to the corresponding
two-component spinor in Eq.\ (\ref{eq:wave}) that is not normalized
for $\phi_0 > \phi_c$. Unlike the probability current discussed
below, the probability density $|\psi|^2$ is not conserved upon
reflection. Indeed (note $\cos\phi_0 = - \cos\phi_1$)
\begin{equation}
  \label{eq:jn}
  |A_0|^2 \cos\phi_0 + |A_1^\pm|^2 \cos\phi_1
  + |A_2^\mp|^2 \cos(\Re \,\phi_2^\mp) = 0,
\end{equation}
which illustrates the importance of $\epsilon_\mp$ for our problem.

The Hamiltonian (\ref{eq:ham}) yields the following expression for
the velocity operator
\begin{equation}
  \label{eq:veldef}
  \vek{v} = \frac{i}{\hbar} [H, \vek{r}]
  = \frac{1}{\hbar} (\mu \kk + \alpha \hat{\vek{e}}_z \times \vek{\sigma}),
\end{equation}
where $\hat{\vek{e}}_z$ denotes a unit vector perpendicular to the
2D plane. We see here that SOC gives rise to a nontrivial
spin-dependent difference between the electrons' momentum and
velocity that plays a crucial role in our analysis below of the spin
angular impulse. Outside the region where the beams interfere and
for real angles $\phi$ we get for the magnitude $v$ of the velocity
\begin{equation}
  \label{eq:vel}
  v \equiv \expect{v} = \frac{1}{\hbar} (\mu k_+ + \alpha)
  = \frac{1}{\hbar} (\mu k_- - \alpha),
\end{equation}
i.e., all beams have the same velocity $v$ (parallel to the
corresponding wave vector). \cite{zue01} For complex angles
$\phi_2^+$ the velocity is slightly larger than Eq.\ (\ref{eq:vel}),
and it is oriented perfectly parallel to the barrier,
\begin{equation}
  \label{eq:velover}
  \vek{v}_2^+ = \frac{\hat{\vek{e}}_y}{\hbar}
  \left[\mu k_y^- + \alpha \:\frac{\sin\phi_c}{\sin\phi_0}\right].
\end{equation}

Similar to Eq.\ (\ref{eq:veldef}), we get for the probability
current
\begin{equation}
  \vek{j} \equiv \expect{\vek{j}} = \frac{1}{\hbar} \big[ \mu\,
  \Re(\langle\psi| \kk | \psi\rangle)
  + \alpha \langle\psi| \hat{\vek{e}}_z \times \vek{\sigma} |\psi\rangle\big].
\end{equation}
We emphasize that unlike $\vek{v} = \expect{\vek{v}}$ and
$\expect{\vek{\sigma}}$, the expectation value $\vek{j} =
\expect{\vek{j}}$ is \emph{not} normalized with respect to the
corresponding wave function. Obviously, this is necessary to obtain
the continuity equation
\begin{equation}
\label{eq:divj}
  \partial_t \rho + \nabla \cdot \vek{j} = 0,
\end{equation}
where $\rho = |\psi|^2$ is the probability density. Of course, in
our case $\partial_t \rho = 0$. For the region where both the
incoming and the reflected beams are present we get $j_x = 0$ (as
expected for an impenetrable barrier). On the other hand, the
current component $j_y$ in this region depends in an oscillatory
fashion on the distance $|x|$ to the barrier due to the interference
of the three terms in Eq.\ (\ref{eq:wave}). We do not give here the
lengthy expressions.

Outside the region where both the incoming and the reflected beams
are present, we get in the undercritical regime [using Eqs.\
(\ref{eq:coeff}) and (\ref{eq:vel})]
\begin{subequations}
  \label{eq:current}
  \begin{eqnarray}
  \vek{j}_0^\pm &  = & |A_0|^2 \, \vek{v}_0, \\
  \vek{j}_1^\pm &  = & |A_0|^2 \,
  \frac{\sin^2 (\phi_0 - \epsilon_\mp/2)}{\cos^2(\epsilon_\mp/2)}
  \:\vek{v}_1, \\
  \vek{j}_2^\mp & = & |A_0|^2 \,
  \frac{\cos^2 \phi_0}{\cos^2(\epsilon_\mp/2)}
  \:\vek{v}_2^\mp,
\end{eqnarray}
\end{subequations}
where $v_i = v$. In the overcritical regime we have
\begin{equation}
  \label{eq:curover}
  \vek{j}_0^- = |A_0|^2 \vek{v}_0, \qquad
  \vek{j}_1^- = |A_0|^2 \vek{v}_1
\end{equation}
with $v_0 = v_1 = v$. For the extraordinarily reflected beam we get
\begin{equation}
  \label{eq:curovere}
  \vek{j}_2^+ (x) = |A_0|^2 \: \frac{2 \cos^2 \phi_0}{1 + \sin\phi_c}
  e^{2\kappa_2^+ x} \:\vek{v}_2^+,
\end{equation}
i.e., the current $\vek{j}_2^+$ dies off exponentially with
increasing distance $|x|$ from the barrier.

We evaluate the currents reflected from a unit segment of the
barrier to get the reflection coefficients
\begin{equation}
  \label{eq:reflect}
    R_{\pm\pm} =
    \frac{\sin^2 (\phi_0 - \epsilon_\mp/2)}{\cos^2(\epsilon_\mp/2)},
    \quad
    R_{\pm\mp} =
    \frac{\cos (\phi_0 - \epsilon_\mp) \cos \phi_0}{\cos^2(\epsilon_\mp/2)},
\end{equation}
where the first (second) sign of $R_{\pm\pm}$ corresponds to the
incoming (reflected) beam [Fig.\ \ref{fig:results}(b)]. Current
conservation implies $R_{++} + R_{+-} = R_{--} + R_{-+} = 1$, which
is equivalent to Eq.\ (\ref{eq:jn}) because $v_i = v$. We note that
in the overcritical regime we have $R_{-+} = 0$. At a first glance
this appears counterintuitive because the reflected current
$\vek{j}_2^+$ is nonzero. However, this current is oriented parallel
to the barrier so that it does not enter the reflection coefficient.

It is known for the Rashba model (\ref{eq:ham}) that propagating
beams are characterized by a spin orientation in the 2D plane and
perpendicular to the corresponding wave vector, \cite{and92, win03}
i.e., for a wave vector $\kk_\pm$ with polar angle $\phi$ the
orientation of the unit vector $\expect{\vek{\sigma}}_\pm$ is
characterized by the angle $\phi \mp \pi/2$. In the overcritical
regime $\phi_0 > \phi_c$, we obtain the out-of-plane spin
orientation \cite{usa05}
\begin{equation}
  \label{eq:spinplus}
  \expect{\vek{\sigma}}_2^+
  = \marr{\;\cos \eta\; \\[0.5ex] 0 \\[0.5ex] \sin\eta},
\end{equation}
where $\eta = \arccos (\sin \phi_c / \sin \phi_0)$. The largest
value of $\eta$ is obtained in the limit of grazing incidence
($\phi_0 \rightarrow \pi/2$) giving $\eta_\mathrm{max} =
|\epsilon_\mathrm{max}|$ [Fig.\ \ref{fig:results}(a)] with
$\eta_\mathrm{max} \rightarrow \pi/2$ for low densities $N
\rightarrow N_q$. 

It is well known that during the elastic reflection of electrons off
an impenetrable barrier, the barrier exerts a force $\vek{F}$ on the
electrons. Yet for such a scattering process only the linear
impulse, i.e., $\vek{F}$ integrated over the time $\Delta t$ of the
collision process is physically meaningful. Obviously, this linear
impulse per electron equals the change $\hbar\Delta \kk$ of crystal
momentum. This result gets modified by the presence of SOC. Using
Eq.\ (\ref{eq:veldef}) we get
\begin{equation}
  \label{eq:impdef}
  \vek{F} \Delta t = m^\ast \left( \Delta \expect{\vek{v}}
    - \frac{\alpha}{\hbar} \hat{\vek{e}}_z \times \Delta
    \expect{\vek{\sigma}} \right)
  =\hbar \Delta{\vek{k}}.
\end{equation}
Furthermore, SOC gives rise to multiple reflected beams as discussed
above. When taking into account the conservation of the electron
number during the scattering process, one finds using a continuous
media approach that $\hbar\Delta \kk$ for the components of the
reflected beam must be weighted by the corresponding reflection
coefficients (\ref{eq:reflect}). 

Equation (\ref{eq:impdef}) implies that the barrier also exerts an
orbital torque that changes the orbital angular momentum of the
electrons. In a similar way (while there is no direct effect of the
barrier on the electron's spin), SOC and the barrier exert an
effective spin torque on the electrons that changes the spin
orientation when the electrons are reflected at the barrier. Using
Eq.\ (\ref{eq:impdef}) we can write the dimensionless spin angular
impulse as
\begin{equation}
  \label{eq:spinimpdef}
  \Delta \expect{\vek{\sigma}}
  = \frac{\hbar}{\alpha} \, \hat{\vek{e}}_z \times
  \left(\frac{\hbar\Delta \vek{k}}{m^\ast}
    - \Delta \expect{\vek{v}} \right),
\end{equation}
which shows that the change of the spin orientation is a combined
effect of SOC and the change in orbital motion characterized by a
nontrivial difference between the changes of the electron's momentum
and velocity. When averaging over the components of the reflected
beam we get
\begin{equation}
  \label{eq:imp}
  \Delta \expect{\vek{\sigma}}^\pm =
    R_{\pm\pm} \left(\expect{\vek{\sigma}}_1^\pm
                   - \expect{\vek{\sigma}}_0^\pm\right)
  + R_{\pm\mp} \left(\expect{\vek{\sigma}}_2^\mp
                   - \expect{\vek{\sigma}}_0^\pm\right).
\end{equation}
$\Delta\expect{\sigma}^\pm$ approaches magnitude $\sim 1$ around
$\phi_0 \simeq \pi/4$ when \emph{on average} the spin orientation of
the electrons becomes zero upon reflection. In other words, the spin
angular momentum carried by a $+$ or $-$ polarized current is fully
absorbed by the barrier around $\phi_0 \simeq \pi/4$. Clearly, this
has important consequences for spin-dependent transport in confined
geometries. \cite{zut04} Also, it offers interesting perspectives
for current-driven domain wall motion and magnetization reversal.
\cite{ral08z} Even for the electrons in an unpolarized incoming beam
the average spin angular impulse
\begin{equation}
  \label{eq:avimp}
  \Delta \expect{\vek{\sigma}} = {\textstyle\frac{1}{2}} \left(
    \Delta \expect{\vek{\sigma}}^+ + \Delta \expect{\vek{\sigma}}^- \right)
\end{equation}
is nonzero. Figure \ref{fig:results}(c) shows that $\Delta
\expect{\sigma}$ can be quite significant and that it is the largest
in magnitude at the critical angle $\phi_c$. For the parameters of
Fig.~\ref{fig:results}, the maximum of $\Delta \expect{\sigma}$
amounts to $0.09$. We note that while $\expect{\vek{\sigma}}_2^+$ in
the overcritical regime is out-of-plane, \cite{usa05} Eq.\
(\ref{eq:spinplus}) does not give rise to an out-of-plane component
of $\Delta \expect{\vek{\sigma}}$ because in this regime we have
$R_{-+}=0$.

Finally, we comment on how our findings depend on the sample
geometry. The SOC in Eq.\ (\ref{eq:ham}) can be interpreted as a
Zeeman term with an effective magnetic field $\vek{\omega} (\kk) =
(2\alpha/\hbar) (k_y,-k_x)$ giving rise to a precessional motion
with frequency $\omega = |\vek{\omega} (\kk)|$. Quite generally, the
deflection of electron trajectories in confined geometries implies
that the orientation of $\vek{\omega} (\kk)$ changes along these
trajectories. If we approximate the deflection by a circular orbit
with radius $R$, we can distinguish two regimes. \cite{win03, fru04}
If $\omega \gg \Omega \equiv \hbar k/(m^\ast R)$, the electron spins
follow adiabatically $\vek{\omega} (\kk)$. Here, $\Delta
\expect{\vek{\sigma}}$ is simply given by the change of
$\vek{\omega} (\kk)$. Thus $\Delta \expect{\vek{\sigma}} = 0$ for an
unpolarized incoming beam. If, on the other hand, $\omega \lesssim
\Omega$, i.e., $R \lesssim R_0 \equiv \hbar^2 / (2m^\ast \alpha)$,
we are in the nonadiabatic regime, where spin eigenstates are
scattered into a superposition of oppositely oriented eigenstates.
(The above discussion corresponds to the limiting case $R=0$.) For
the parameters used in Fig.~\ref{fig:results}, we have $R_0 =
270$~nm. In systems with weaker SOC than InSb, $R_0$ is yet larger.
Therefore, taking typical sample dimensions into account, the spin
angular impulse discussed here is important for a large variety of
spin-dependent transport experiments in confined geometries.
\cite{zut04} We note that spin relaxation lengths are usually
significantly larger than~$R_0$.

In conclusion, our analysis demonstrates that the spin-dependent
reflection provides a new mechanism that changes the spin
orientation via the spin angular impulse exerted on the electrons
when they are reflected off a barrier in the presence of SOC. While
the present work has focused for conceptual clarity on a straight
and infinitely high barrier, the underlying physics is relevant for
a large variety of transport experiments in confined geometries
including soft barriers or sample boundaries with different shapes.
The mechanism provides interesting possibilities for current-driven
magnetization dynamics.
RW appreciates stimulating discussions with J.~Heremans and
U.~Z\"ulicke. Work at Argonne was supported by DOE BES under
Contract No.\ DE-AC02-06CH11357.


\begin{thebibliography}{10}
\bibitem{quantdens}
For the parameters in Fig.~\ref{fig:results}, we have $N_q = 1\times
  10^8$~cm$^{-2}$. For $N < N_q$ the incoming beam has two different wave
  vectors $\kk_-$. Then we are always in the overcritical regime discussed
  below.
\end{thebibliography}
\end{document}